\documentclass[%
  superscriptaddress,
  reprint, twocolumn,
  amsmath,amssymb,
  aps,prb,
  floatfix,
]{revtex4-2}

\usepackage{graphicx}      %
\usepackage{dcolumn}       %
\usepackage{bm}            %

\usepackage[dvipsnames]{xcolor}

\usepackage{hyperref}
\hypersetup{
  colorlinks=true,
  citecolor=blue,
  linkcolor=blue,
  filecolor=blue,
  urlcolor=blue,
  pdfstartview=FitH,
  pdfpagemode=UseNone
}

\usepackage{svg}
\usepackage{placeins}
\usepackage{booktabs}
\usepackage{soul}

\usepackage{multirow}
\usepackage{amsfonts}
\usepackage{amsthm}
\usepackage{mathrsfs}
\usepackage[title]{appendix}
\usepackage{textcomp}
\usepackage{algorithm}
\usepackage{algorithmicx}
\usepackage{algpseudocode}
\usepackage{listings}

\usepackage[T1]{fontenc}

\newcommand{\orcid}[1]{%
  \href{https://orcid.org/#1}{\includegraphics[width=8pt]{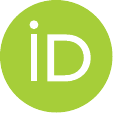}}}

\begin{document}

\title[Fuzzy NN for Quantum Wavefunction]{Fuzzy Neural Network Performance and Interpretability of \\ Quantum Wavefunction Probability Predictions}

\author{Pedro H. M. Zanineli \orcid{0009-0008-2359-5218}} 
\email{pedro.zanineli@lnnano.cnpem.br}
\affiliation{Brazilian Nanotechnology National Laboratory (LNNano/CNPEM), 13083-100, Campinas, SP, Brazil}
\affiliation{Universidade Federal do ABC (UFABC), 09210-580, Santo André, São Paulo, Brazil}

\author{Matheus Zaia Monteiro}
\email{matheus.z.monteiro@gmail.com}
\affiliation{Ilum School of Science, Brazilian Center for Research in Energy and Materials, 13087-548, Campinas, SP, Brazil}

\author{Vinicius Francisco Wasques}
\email{vinicius.wasques@ilum.cnpem.br}
\affiliation{Ilum School of Science, Brazilian Center for Research in Energy and Materials, 13087-548, Campinas, SP, Brazil}

\author{Francielle Santo Pedro Simões}
\email{fran.stopedro@gmail.com}
\affiliation{Multidisciplinary Department, Federal University of São Paulo, 06110-295, Osasco, SP, Brazil}

\author{Gabriel R. Schleder \orcid{0000-0003-3129-8682}}
\email{gabriel.schleder@lnnano.cnpem.br}
\affiliation{Brazilian Nanotechnology National Laboratory (LNNano/CNPEM), 13083-100, Campinas, SP, Brazil}
\affiliation{Universidade Federal do ABC (UFABC), 09210-580, Santo André, São Paulo, Brazil}

\begin{abstract}
Predicting quantum wavefunction probability distributions is crucial for computational chemistry and materials science, yet machine learning (ML) models often face a trade-off between accuracy and interpretability. 
This study compares Artificial Neural Networks (ANNs) and Adaptive Neuro-Fuzzy Inference Systems (ANFIS) in modeling quantum probability distributions for the H$_{2}^+$ ion, leveraging data generated via Physics-Informed Neural Networks (PINNs). 
While ANN achieved superior accuracy (R$^2$ = 0.99 vs. ANFIS’s 0.95 with Gaussian membership functions), it required over 50$\times$ more parameters (2,305 vs. 39–45). 
ANFIS, however, provided unique interpretability: its Gaussian membership functions encoded spatial electron localization near proton positions ($\mu = \pm 1.2\ \text{\AA}$), mirroring Born probability densities, while fuzzy rules %
reflected quantum superposition principles. 
Rules prioritizing the internuclear direction revealed the system’s 1D symmetry, aligning with Linear Combination of Atomic Orbitals theory—a novel data-driven perspective on orbital hybridization.
Membership function variances ($\sigma$) further quantified electron delocalization trends, and peak prediction errors highlighted unresolved quantum cusps. 
The choice of functions critically impacted performance: Gaussian/Generalized Bell outperformed Sigmoid, with errors improving as training data increased, underscoring scalability. 
This study underscores the context-dependent value of ML: ANN for precision and ANFIS for interpretable, parameter-efficient approximations that link inputs to physical behavior. 
These findings advocate hybrid approaches in quantum simulations, balancing accuracy with explainability to accelerate discovery. Future work should extend ANFIS to multi-electron systems and integrate domain-specific constraints (e.g., kinetic energy terms), bridging data-driven models and fundamental physics.  
\end{abstract}


\maketitle

\section{Introduction}\label{sec1}

In the framework of quantum mechanics, the wavefunction $\Psi$ represents a fundamental concept, encapsulating the state of an isolated quantum system. This function assigns a complex value to every point in space, generating a probability amplitude that, when interpreted through the Born rule, translates into actual charge densities or probability distributions \cite{griffiths_2018}.

A crucial property of $\Psi$ is that its squared modulus, $|\Psi|^2$, provides the probability density of finding a particle at a specific location. A well-known example is the hydrogen atom, where a single-electron wavefunction describes the likelihood of detecting the electron within a given spatial region \cite{griffiths_2018}.

Importantly, determining the wavefunction of a system does not only offer insight into spatial probability distributions but also provides a comprehensive description, including observables quantities such as energy and momentum obtained via their corresponding operators. The wavefunction is obtained by solving the Schr\"odinger equation:
$$\hat{H} \Psi = E \Psi,$$
where the Hamiltonian operator $\hat{H} = \left[ -\frac{\hbar^2}{2m} \nabla^2 + V(\mathbf{r}) \right]$ dictates the system's total energy, with its eigenvalues corresponding to possible energy levels and eigenvectors representing the associated quantum states. However, solving this equation is a significant challenge, particularly for many-body or complex systems \cite{computational_2011,DFT_review}.

Analytical solutions exist solely for a limited number of scenarios, including a free particle, the particle in a box, the quantum harmonic oscillator, and the hydrogen atom \cite{griffiths_2018,MLIPs2024}. Thus, different approaches of machine learning (ML) have been used for finding the equation's solution, such as neural networks \cite{nn_2020}, generative models \cite{generative_2023}, genetic algorithms \cite{GA_2001}, and reinforcement learning \cite{reinforcement_2019}. 

Due to these limitations, alternative approaches such as Physics Informed Neural Networks (PINNs) have been explored to enhance our ability to find approximate solutions for more intricate quantum systems \cite{pinn_eigenvalue_2022,advances_2023,prl_pinn_2024}. In particular, PINNs have been utilized to determine the solutions for the hydrogen molecular ion $\text{H}_2^+$ \cite{physicsinformed_2022}.

By leveraging fundamental physical principles, this method identifies parametric eigenvalues and eigenfunction surfaces of quantum systems. Notably, this approach is capable of enabling \textit{ab initio} calculations for simple molecular structures. Considering the importance of uncovering the probability density of a wavefunction, creating a machine learning model from the PINN-generated data can bring advantages for different reasons.

Firstly, knowing that solving the Schr\"odinger equation is computationally costly, using ML would be interesting for finding an approximate solution with less time \cite{ML_eq_2020}. Secondly, if the ML is well-suited, it can be generalized for different systems, which can be used for different molecules and materials. Finally, according to the chosen model, it can also be important from the perspective of interpretability and physical analysis.

In this last case, while on one hand classical Artificial Neural Networks (ANN) are known for hindering the result's interpretability \cite{interp_2021}, on the other hand, Adaptive Neuro-Fuzzy Inference System (ANFIS) might be suitable for better comprehending the pattern found from the neural network \cite{anfis_2018,anfis_vs_ann_2016}. Yet, it is fundamental to find a balance between the model's efficiency and interpretability.

Thus, in the light of the exposed, the present work aims to establish a comparison between classical Artificial Neural Networks and the Adaptive Neuro-Fuzzy Inference System in the wavefunction prediction of a dihydrogen ion.

\section{Some preliminaries}

This section provides some mathematical tools to better understand how ANFIS works. 

\subsection{Fuzzy Sets and Fuzzy Logic}
\label{subsec:fuzzysetsfuzzylogic}

Fuzzy sets extend the concept of classical sets by introducing a membership function $\varphi:U\to [0, 1]$, which assigns each element of the universe $U$ to a value within the interval $[0,1]$. In other words, considering a subset $A$ in the universe $U$, the value $\varphi_A(x)$ means the degree of association of the element $x$ to the set $A$, in the following sense, the higher the degree, the higher is the association. This generalization enables the representation of vague or imprecise relationships that are common in natural language but lack precise mathematical definitions. Fuzzy sets address this issue by allowing elements to belong to a set with varying degrees of membership, rather than the strict binary classification of classical sets \cite{barros2017}.  

Fuzzy logic expands classical logic by redefining logical operators such as conjunctions and disjunctions through $t$-norms and $t$-conorms (or also called $s$-norms), respectively. A $t$-norm is a function $t:[0,1]^2\to[0,1]$ that satisfies the properties of commutativity, monotonicity, and associativity. Examples of $t$-norms include the minimum and the algebraic product functions. Also, $t$-conorms satisfy similar properties, and they are connected by the following relation $s(x,y) = 1 - t(1-x,1-y)$, where $t(\cdot, \cdot)$ represents a $t$-norm and $s(\cdot, \cdot)$ a $t$-conorm. Common examples of $t$-conorms include the maximum operator and the algebraic sum \cite{klir1995}.

Fuzzy logic introduces degrees of truth rather than the traditional binary true or false statements. Using fuzzy sets, fuzzy propositions take the form:
\begin{quotation}
\centering
    ``IF \textit{antecedent}, THEN \textit{consequent}'',
\end{quotation}
where both \textit{antecedent} and \textit{consequent} are represented by fuzzy sets. A typical fuzzy rule can be expressed as:
\begin{quotation}
\centering
    ``IF $x_1$ is $A_1$ AND $x_2$ is $A_2$, THEN $y$ is $B$'',
\end{quotation}
where $A_1$, $A_2$, and $B$ are fuzzy sets defined over the respective domains of $x_1$, $x_2$, and $y$, and the AND operator is modeled by a $t$-norm. A fuzzy proposition can be interpreted as a function $P:U_1\times U_2\to [0, 1]$, where the codomain represents the degree of truth of the proposition.

\subsection{Fuzzy Rule-Based Systems (FRBS)}
\label{subsec:frbs}

Fuzzy Rule-Based Systems (FRBSs) employ fuzzy propositions to replicate human decision-making. The system begins by mapping numerical inputs onto fuzzy sets in a process known as fuzzification. Thus, it applies an inference system to combine multiple ``IF-THEN'' fuzzy rules, aggregating them into an overall decision. If the conclusions of these rules are fuzzy sets, a defuzzification step converts them into numerical outputs.

Several inference strategies are used in FRBSs. One of the most well-known is the Mamdani method, which produces fuzzy sets as outputs that require defuzzification to yield numerical values. In this approach, the AND operator is typically implemented as the minimum function, and the activation weight of the antecedent is computed as:
\begin{equation*}
    \varphi_{\mathcal{R}_i}(x_1, \dots, x_n, y) = \min\left[\varphi_{A_1^{(i)}}(x_1), \dots, \varphi_{A_n^{(i)}}(x_n), \varphi_{B^{(i)}}(y) \right],
\end{equation*}
where $A_j^{(i)}$ represents the fuzzy set for the $i$-th rule corresponding to the $j$-th input variable in an FRBS with $n$ inputs and one output. The final fuzzy set is determined by aggregating individual rule outputs using the maximum function:
\begin{align*}
    \varphi_{\mathcal{M}}(x_1, \dots, x_n, y) &=\\ &\max\left[\varphi_{\mathcal{R}_1}(x_1, \dots, x_n, y), \dots, \varphi_{\mathcal{R}_n}(x_1, \dots, x_n, y)\right].
\end{align*} 

A generalization of this inference method, with multiple outputs can be found in \cite{gomide2007}.

Another commonly used inference method is the Takagi-Sugeno-Kang (TSK) model, which expresses rule consequents as mathematical functions, typically linear combinations of input variables. Using the algebraic product as the AND operator, each rule yields an output computed as:
\begin{equation*}
    y^{(i)} = b^{(i)} + \sum_{j=0}^{n}a_j^{(i)}x_j,
\end{equation*}
where $b^{(i)}$ and $a_j^{(i)}$ are coefficients associated with the $i$-th rule. The final system output is obtained via:
\begin{equation*}
    \hat y = \frac{\sum_i w_iy^{(i)}}{\sum_i w_i},
\end{equation*}
where $w_i$ represents the activation weight of the $i$-th rule.

\subsection{Adaptive Neuro-Fuzzy Inference System (ANFIS)}
\label{subsec:anfis}

\begin{figure}[!h]
    \centering
    \includegraphics[width=1\linewidth]{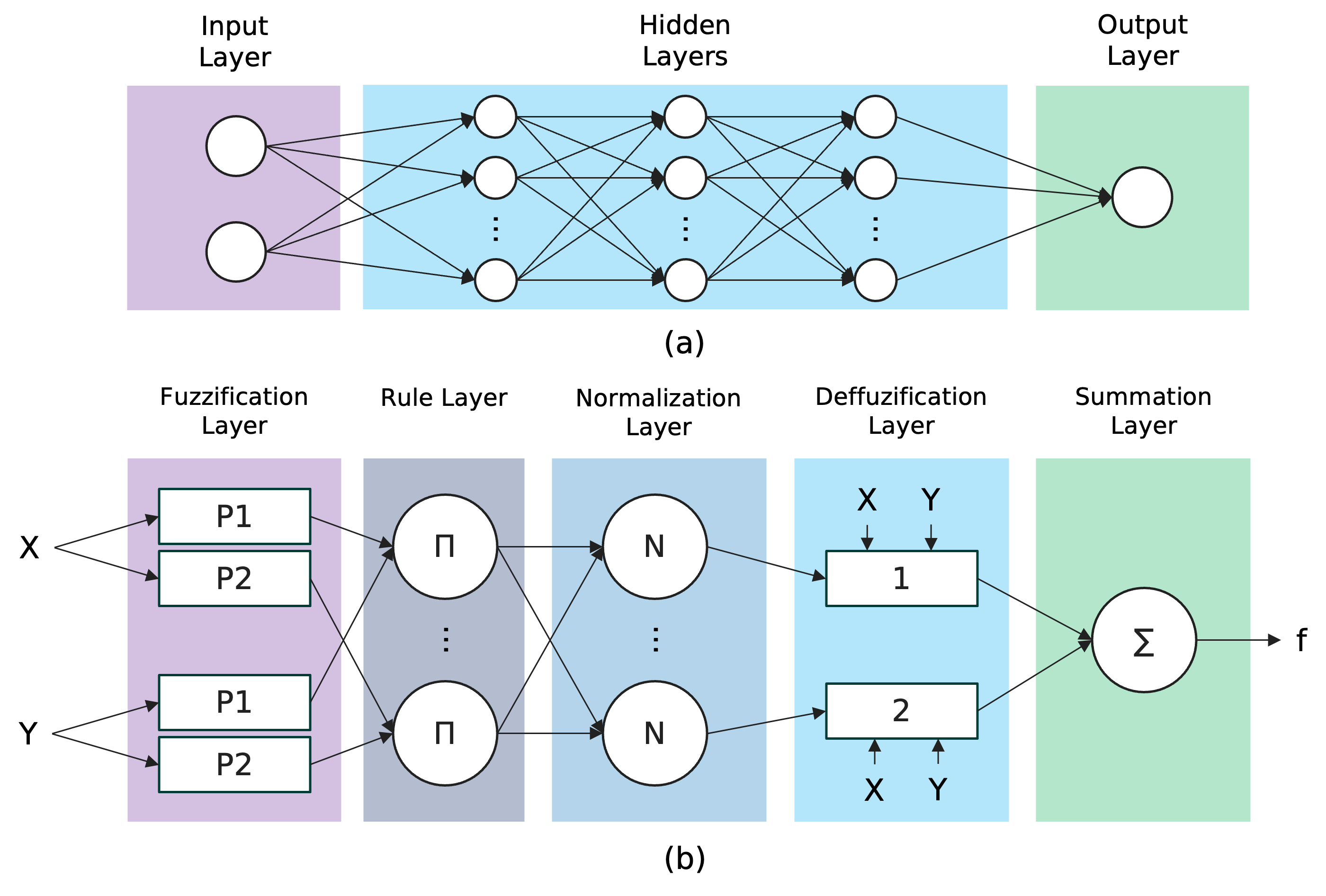}
    \caption{Artificial Neural Network and Adaptive Neuro-Fuzzy Inference System architectures in (a) and (b), respectively.}
    \label{fig:anfis}
\end{figure}

Deep Learning techniques focus on capturing patterns, they lack interpretability, making it difficult to understand how predictions are derived. The Adaptive Neuro-Fuzzy Inference System (ANFIS)  bridges this gap by integrating fuzzy logic with neural networks. ANFIS combines a TSK-based FRBS with backpropagation, enabling it to learn both fuzzy set parameters and rule consequents. Figure \ref{fig:anfis} presents an ANFIS structure with two input variables \cite{jang1993}.

\section{Methodology}

In an initial moment, the Physics-Informed Neural Network (PINN) model for generating the data was used with a single interatomic distance between the dihydrogen ion.
Therefore, the PINN solves for the eigenvalues and eigenfunctions of the Hamiltonian operator \cite{physicsinformed_2022}:
\begin{align}
\newcommand{\rr}{ {\bf r}  }
\newcommand{\RR}{ {\bf R}  }
\newcommand{\HH}{ {\hat {{H}} }}
\label{eq:Hamiltonian}
    \HH = - \frac{1}{2} \nabla^2 -   \frac{1}{| {\rr}-\RR_1|}  -   \frac{1}{| {\rr}-{\RR_2}|},
\end{align}
using atomic units, ${\bf r}=(x,y,z)$, and the molecule oriented along the $x$-axis, so ${\bf R}_1 = -{\bf R}_2 = (R,0,0)$.
In this case, the electronic position along $x$ and $y$ axis in the plane $z=0$ was used as features for the models, whereas the probability was used as the target for the model, with a total of 200 points for the dataset.

This data was further split between train, validation, and test. The size defined for the test was 15\%, and 10\% of the 85\% data for the test was used for validation. The further described training was performed in Python using PyTorch \cite{torch}.

All models were trained for 200 epochs, with ADAM as the optimizer and MSE as the loss function. A scheduler was used to diminish the learning rate by a factor of 10 if the test loss did not increase for 10 epochs. Furthermore, for the ANN, the chosen architecture was a fully connected feedforward neural network with three layers. The number of neurons in each layer was 64, 32, and 1, respectively, with ReLU as the activation function.

For the ANFIS model, the architecture was defined in terms of membership functions and was trained using different types of membership functions (MFs): Generalized Bell, Gaussian, and Sigmoid. The Gaussian MF was trained with a 0.01 learning rate, while the Bell and Sigmoid MFs were trained with a 0.005 learning rate.

An Adaptive-Network-Based Fuzzy Inference System (ANFIS) consists of five layers that combine fuzzy logic and neural networks for modeling complex relationships \cite{jang1993}. The layers include:
\begin{itemize}
    \item Fuzzification Layer: Computes fuzzy membership values for input variables using predefined membership functions.
    \item Rule Layer: Evaluates fuzzy rules by computing their activation levels based on input membership values.
    \item Normalization Layer: Normalizes rule activations to ensure their sum equals one.
    \item Defuzzification Layer: Combines the outputs of the fuzzy rules using a weighted average or centroid method to generate crisp values.
    \item Output Layer: Normalizes the rule activations to ensure their sum equals one, which is typical for Sugeno-type fuzzy systems.
\end{itemize}

The training process for ANFIS involves optimizing both the membership function parameters and the rule parameters using a hybrid learning algorithm that combines backpropagation and least squares estimation.

To further assess the performance of the ANFIS model, additional experiments were conducted by increasing the number of training points to analyze the impact on loss reduction.

The simulations were realized on a workstation system, using an NVIDIA GeForce RTX 4090 GPU with 24GB RAM.

\section{Results}

\begin{figure}[h]
    \centering
    \includegraphics[width=1\linewidth]{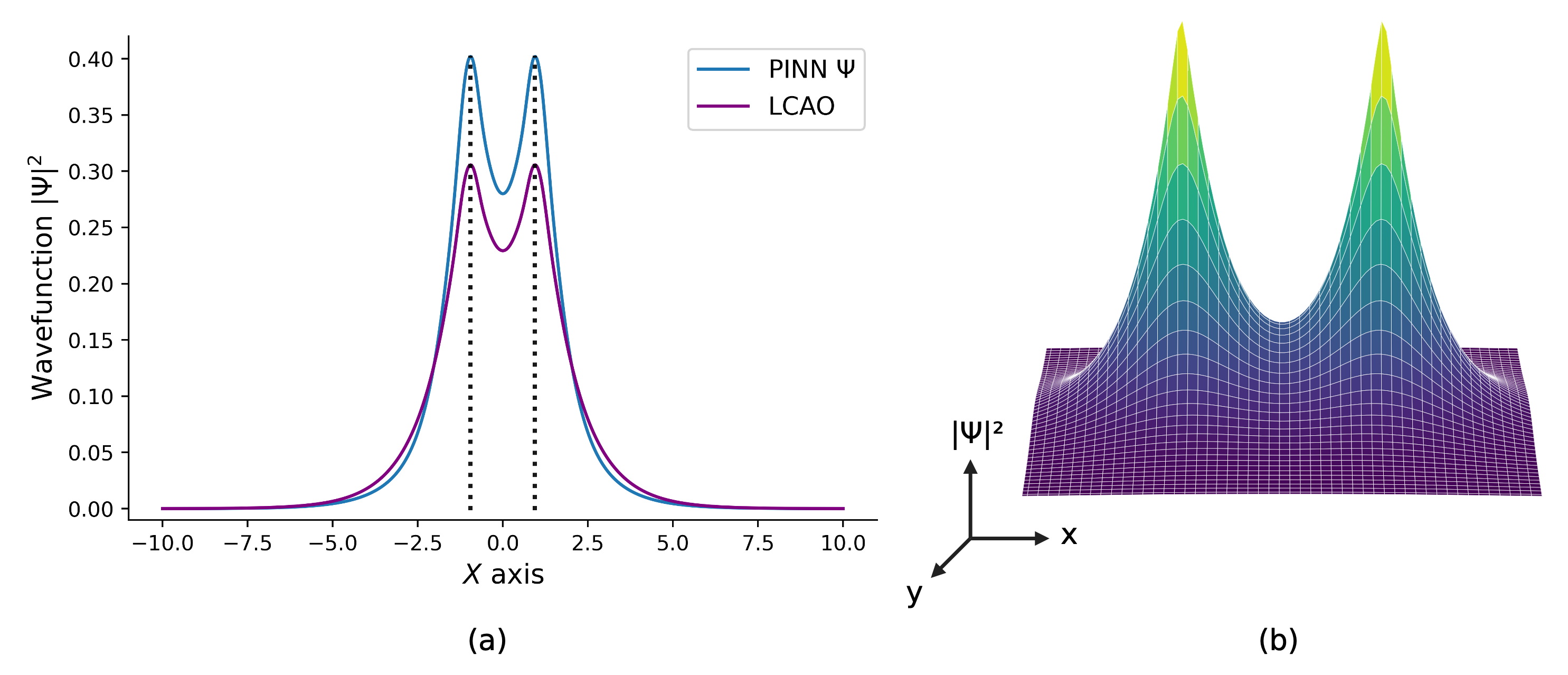}
    \caption{Wavefunction probability amplitude $|\Psi^2|$ of the PINN-generated solution for the H$_2^+$ ion, visualized as (a) a line profile along the $x$-axis (the internuclear direction) and (b) the two spatial dimensions, for the fixed plane $z=0$. In (a), the LCAO solution is presented as a reference comparison, lacking complex interaction terms between the atomic orbitals.}
    \label{fig:data}
\end{figure}

As an initial step, it is possible to visualize the wavefunction data in Figure \ref{fig:data}. In (a), the wavefunction prediction is represented in terms of the interatomic distance for the molecule in the $x$ axis, and in (b) the $x$ and $y$ axis are used to visualize the probability density for the molecule in $z$.

As a result of the usage of the PINN-trained model, the resulting data is shown in Figure \ref{fig:data}. In (a), a cut in the $x$ axis shows the distance between the two atoms, represented by the gap between the black dashed lines, the neural network wavefunction output as the blue line, and the simple linear combination of atomic orbitals (LCAO) solution in purple.

Similarly, in the case of Figure \ref{fig:data} (b), the tridimensional visualization is shown, in which the higher the points, the higher the wavefunction. That is, the greater the values are, the higher is the probability of finding an electron in space.

\begin{figure}[h]
    \centering
    \includegraphics[width=1\linewidth]{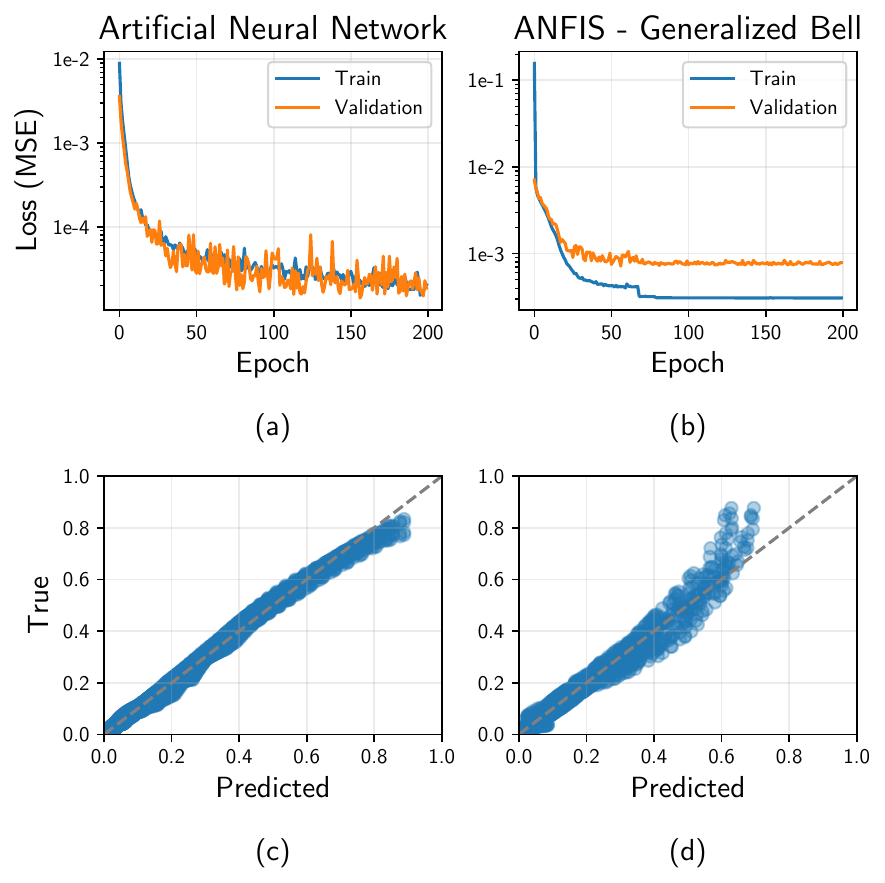}
    \caption{Parity plots comparing the reference and predicted values for the (a) classical and (b) fuzzy methods, respectively, along with their corresponding  models training and validation losses in (c) and (d).}
    \label{fig:initial_comparison}
\end{figure}

Following on with the application of the ANN and ANFIS models, the results can be visualized in Figure \ref{fig:initial_comparison}. As a result, the ANFIS result shows a substantial decrease in the training and validation loss in Figure (d). However, its parity plot evidences that the prediction values do not present a great alignment with the correct values in (b).

Regarding the classical Artificial Neural Network, it is possible to notice in Figure  \ref{fig:initial_comparison}(c) that the training and validation loss rapidly decrease through the epochs, even faster when compared with the ANFIS model. This decrease is followed by a stabilization in a minimum that oscillates in the further epochs. In Figure (a), the parity plot demonstrates a greater alignment of the values.

For a comparison metric, the determination coefficient $R^2$ can be calculated, showing a 0.95 alignment of the ANFIS with Gaussian membership functions against 0.99 for the Artificial Neural Network.

\begin{figure*}
    \centering
    \includegraphics[width=1\linewidth]{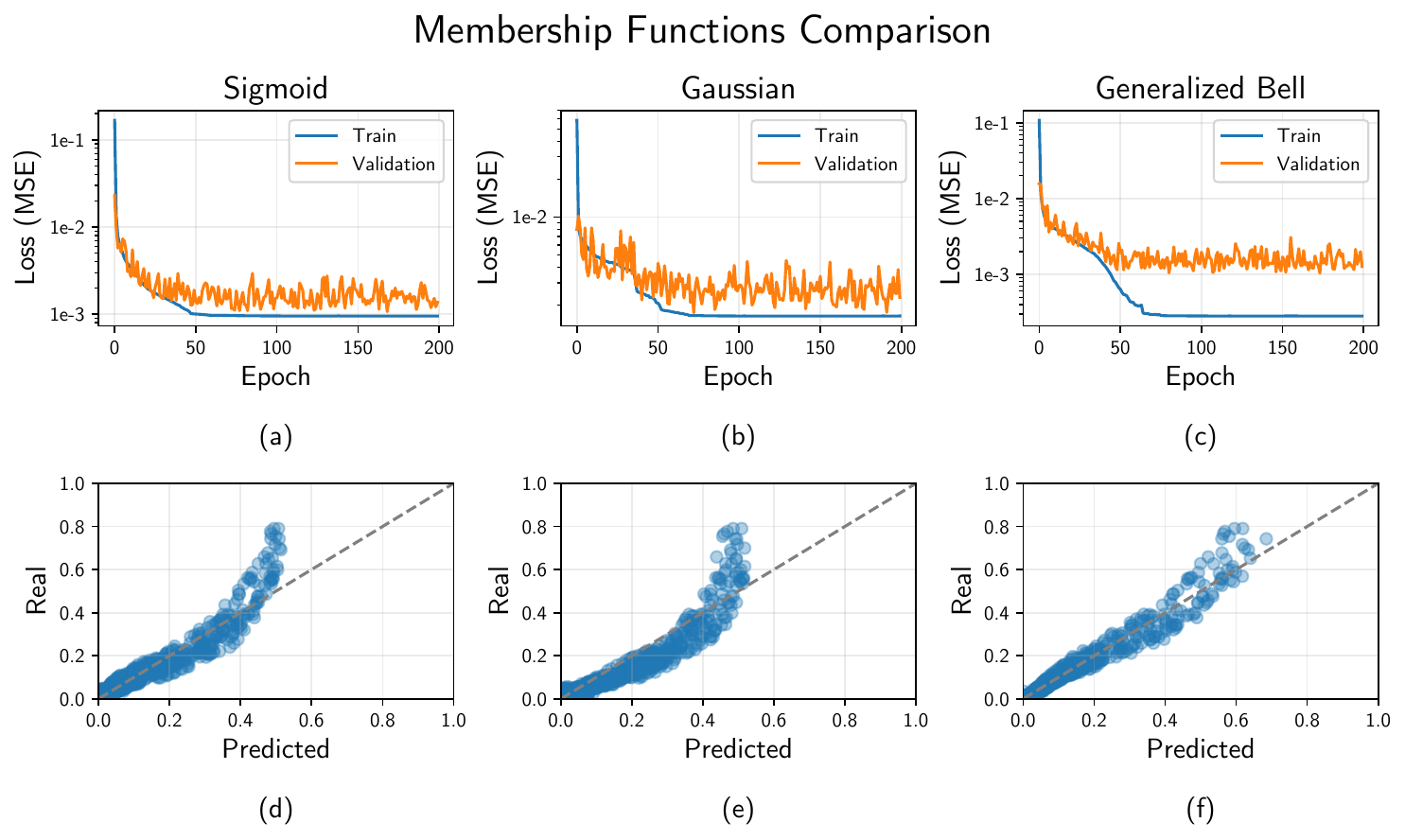}
    \caption{Comparison between the Gaussian, Generalized Bell, and Sigmoid membership function models; in (a-c) are presented the training and validation losses, and in (d-f), the parity plots of the predictions, respectively.}
    \label{fig:different_mfs}
\end{figure*}

If we now test altering the membership functions, we can check the impact of using the Generalized Bell and Sigmoid membership functions. For all three membership functions, the train and validation loss are shown in Figure \ref{fig:different_mfs} (a), (b), and (c), respectively. In particular, it is notable that the Generalized Bell achieves a low loss when compared to the Sigmoid function.

As shown in Figure \ref{fig:different_mfs} (d), (e), and (f), although the Sigmoid membership presents a worse result when compared to the Gaussian in terms of linear alignment of the values, the Generalized Bell showed an equivalent result to the Gaussian. The predicted values are closer to the real values, indicating a better prediction and pattern recognition of the data.

It is curious to note how different membership functions affect the ANFIS training. Even though the probability density generated by the wavefunction presents two tridimensional Gaussian shapes, using a Gaussian and a Generalized Bell membership functions can find great approximated solutions to the problem.

It is also important not to limit the analysis to the determination coefficient, but also base the comparison on the mean absolute error (MAE) and root mean squared error (RMSE), as summarized in Table \ref{tab:metrics}.

As expected from the parity plot of Figure \ref{fig:different_mfs}, the Generalized Bell and Gaussian present similar results. In turn, the classical neural network performs slightly better when compared to the best ANFIS model, which in this case is the Gaussian membership function, with higher scores in two metrics.

Although the classical model presented a better performance, it is fundamental to consider that it used a greater amount of parameters. While the ANFIS model was initialized with 39 and 45 parameters, the ANN had 2035. This is an interesting point for understanding the sensibility of the fuzzy model and better understand the results.

\begin{table}[h]
\caption{Performance comparison between the classical ANN versus ANFIS with different membership functions and number of parameters. For the regression metrics $R^2$, MAE, and RMSE, bold numbers denote the best result.}\label{tab:metrics}%
\resizebox{\linewidth}{!}{%
\begin{tabular}{@{}lccccc@{}}
\toprule
\textbf{Model} & \textbf{Parameters} & \textbf{Grid Size} & \textbf{R$^2$ ($\uparrow$)} & \textbf{RMSE ($\downarrow$)} & \textbf{MAE ($\downarrow$)} \\
        \midrule
        ANN & 2305 & 100 & 0.9956 & 0.008 & 0.0051 \\
            &      & 200 & 0.9991 & 0.0037 & 0.0024 \\
            &      & 300 & \textbf{0.9996} & \textbf{0.0025} & \textbf{0.0017} \\
            &      & 400 & 0.9995 & 0.0027 & 0.0018 \\
        \midrule
        Sigmoid & 39 & 100 & 0.9150 & 0.0379 & 0.0210 \\
                &    & 200 & 0.9643 & 0.0235 & 0.0119 \\
                &    & 300 & 0.9804 & 0.0185 & 0.0078 \\
                &    & 400 & \textbf{0.9856} & \textbf{0.0157} & \textbf{0.0070} \\
        \midrule
        Gaussian & 45 & 100 & 0.9602 & 0.0259 & 0.0145 \\
             &    & 200 & 0.9749 & 0.0197 & 0.0110 \\
             &    & 300 & \textbf{0.9831} & \textbf{0.0171} & \textbf{0.0102} \\
             &    & 400 & 0.9729 & 0.0215 & 0.0126 \\
        \midrule
        Gen. Bell & 39 & 100 & 0.8649 & 0.0478 & 0.0283 \\
              &    & 200 & 0.9753 & 0.0196 & 0.0126 \\
              &    & 300 & 0.9582 & 0.0269 & 0.0129 \\
              &    & 400 & \textbf{0.9921} & \textbf{0.0116} & \textbf{0.0068} \\
\botrule
\end{tabular}
}
\end{table}

\begin{figure}[h]
    \centering
    \includegraphics[width=1\linewidth]{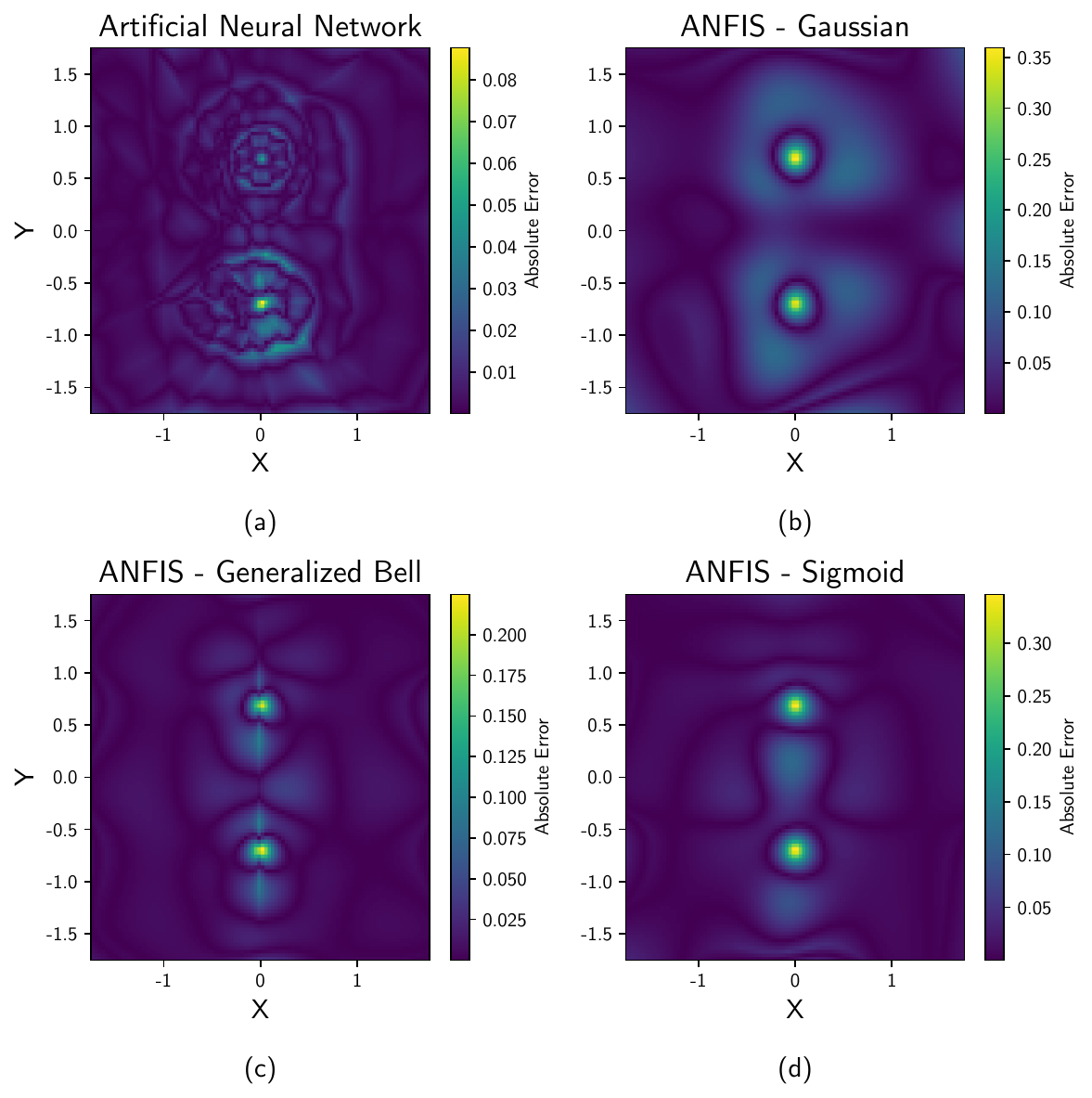}
    \caption{Absolute errors for the probability amplitude in the $xy$-plane for the ANN, Gaussian, Generalized Bell, and Sigmoid models in (a-d), respectively.}
    \label{fig:error}
\end{figure}

Using the four resulting models to predict the probability density in the space, we can then calculate the absolute error and visualize its result, as shown in Figure \ref{fig:error}. Since the present problem is limited to only two independent variables, the visualization is facilitated and can be useful for visualizing better and worse regions of prediction.

Using a color map to indicate the error values, the brighter the color, the greater the error. Thus, regions with darker colors represent regions that were well-predicted, which is the case of the ANN in Figure \ref{fig:error} (a). As expected from the high accuracy, the color map bar indicates low values, representing a prediction that approximates the real solution.

On the other hand, the ANFIS models in Figure \ref{fig:error} (b), (c), and (d) for all MFs have brighter colors, especially in the peak points of the data. This might indicate that the model struggles when predicting the variation in the space. That is, when the regime of probability increases, the model fails at capturing the variation.

Considering the model is being trained with a low volume of data, a possible alternative for this problem is increasing the point count.

Considering that all ANFIS models were trained with a small number of points in the dataset and generated with a low set of parameters, an interesting evaluation is testing the increase of points in the grid. For better comparison between the models, the grids with 100, 200, 300, and 400 points were tested.

\begin{figure}[h]
    \centering
    \includegraphics[width=1\linewidth]{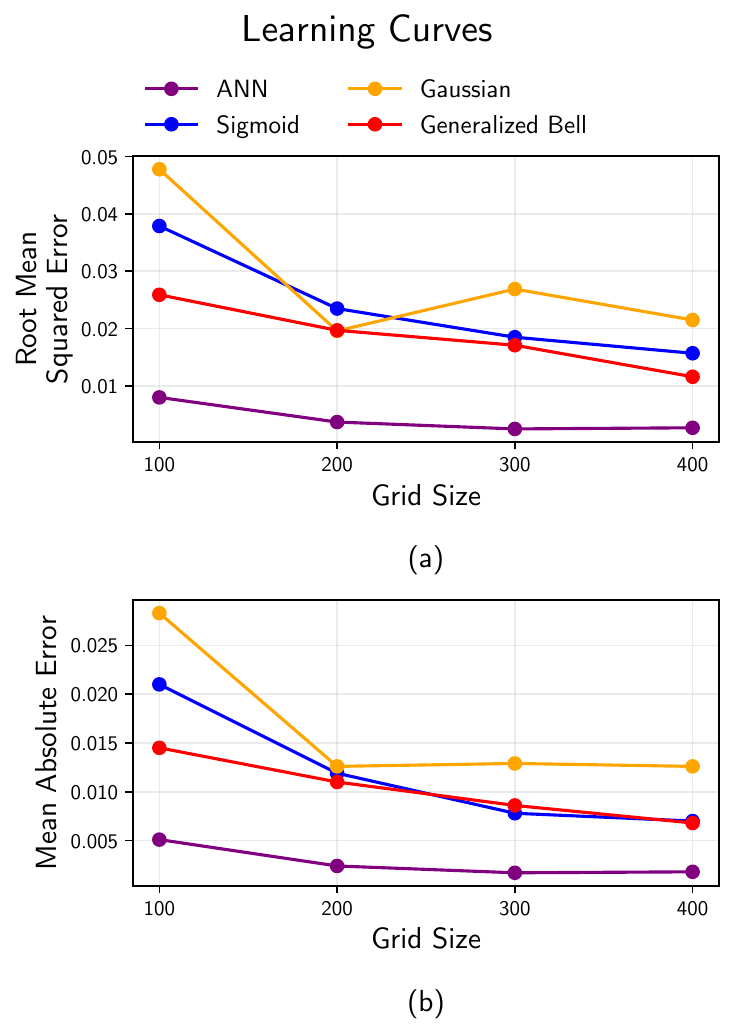}
    \caption{(a) Root mean squared error and (b) mean absolute error for the ANFIS models with different membership functions in terms of the dataset grid size. In blue, yellow, and red are shown the Sigmoid, Gaussian, and Generalized Bell MFs, respectively.}
    \label{fig:loss_grid}
\end{figure}

Comparing all three membership functions as a function of the grid size (Figure \ref{fig:loss_grid}), there is a decrease in model losses for almost all cases of grid size increase. Although the Gaussian MF reaches a plateau in loss decrease, the Sigmoid and Generalized Bell MFs still decrease the mean absolute error in all tests.

Finally, we observe that the Sigmoid and Generalized Bell MFs converge to a similar loss, although the Gaussian model presented an initial better performance. 
Even though the Gaussian MF MAE metric is lower when compared to the other MFs, this model presents results comparable with the ANN with 100 data points, using less than 2\% of parameters, as shown in Table \ref{tab:metrics}.

\begin{figure*}[ht!]
    \centering
    \includegraphics[width=.8\linewidth]{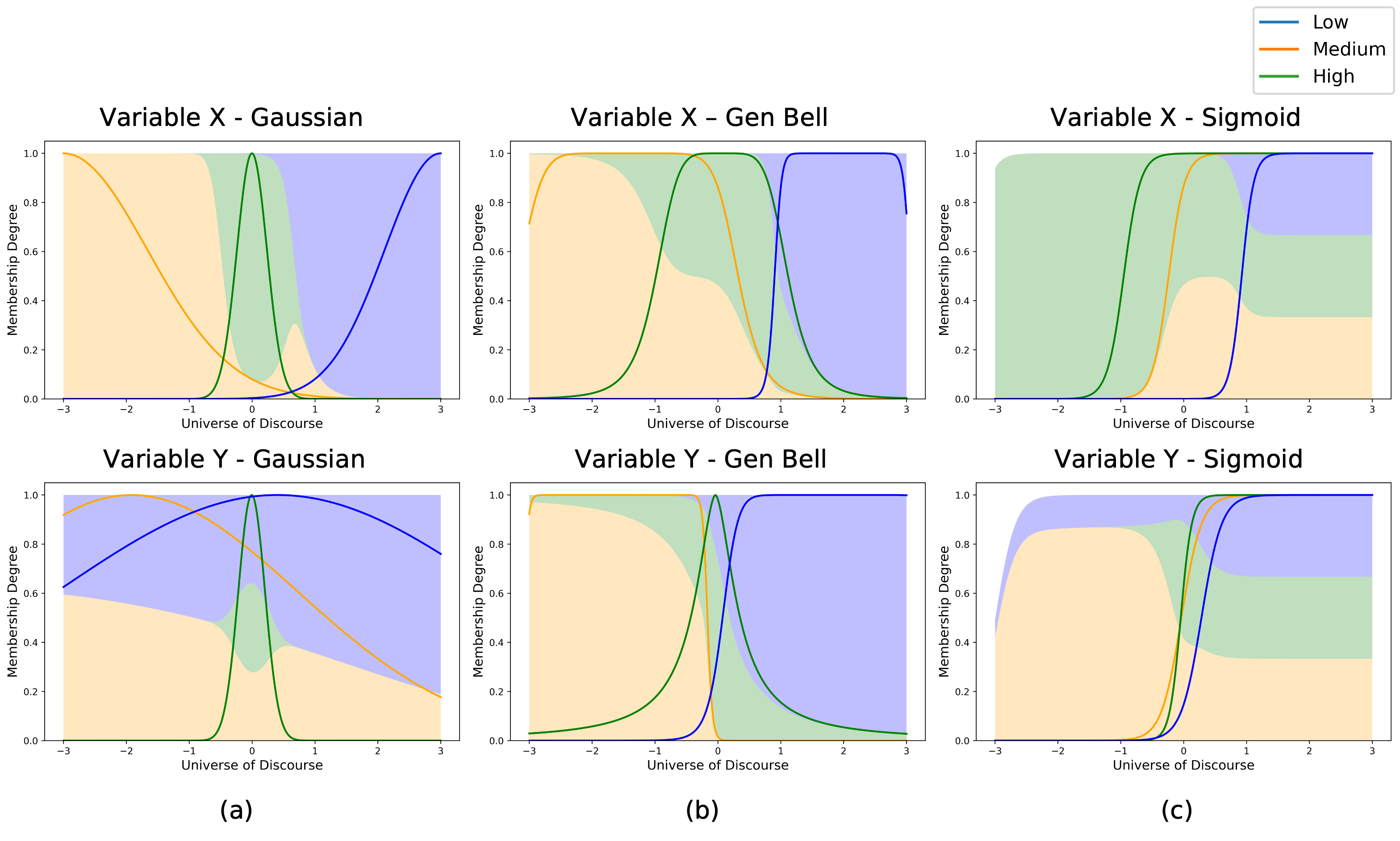}
    \caption{Membership functions for each state variable for the (a) Gaussian, (b) Generalized Bell, and (c) Sigmoid functions, respectively.}
    \label{fig:mfs_plot}
\end{figure*}

Despite the better results for the ANN model, the ANFIS might still allow a better interpretability of the results. As illustrated in Figure \ref{fig:mfs_plot}, the membership functions with their trained parameters are shown in (a), (b), and (c) for the Gaussian, Generalized Bell, and Sigmoid functions, respectively. The filling in the plots represent the variable activation in each rule. 

While on one hand, it is possible to visualize how the Sigmoid membership function faults at capturing the double Gaussian-like format, on the other hand, it is notable how the Gaussian and Generalized Bell better captures it.

As shown in the best two models after training, medium and high functions have the established parameters for both variables that capture the increase of probability through the space, as shown by the increase of the membership degree. In comparison with the Sigmoid membership function, the Gaussian and the Generalized Bell have a wider range in the universe of discourse, which represents a better capture of the whole function domain. 

\begin{figure}[!hb]
    \centering
    \includegraphics[width=\linewidth]{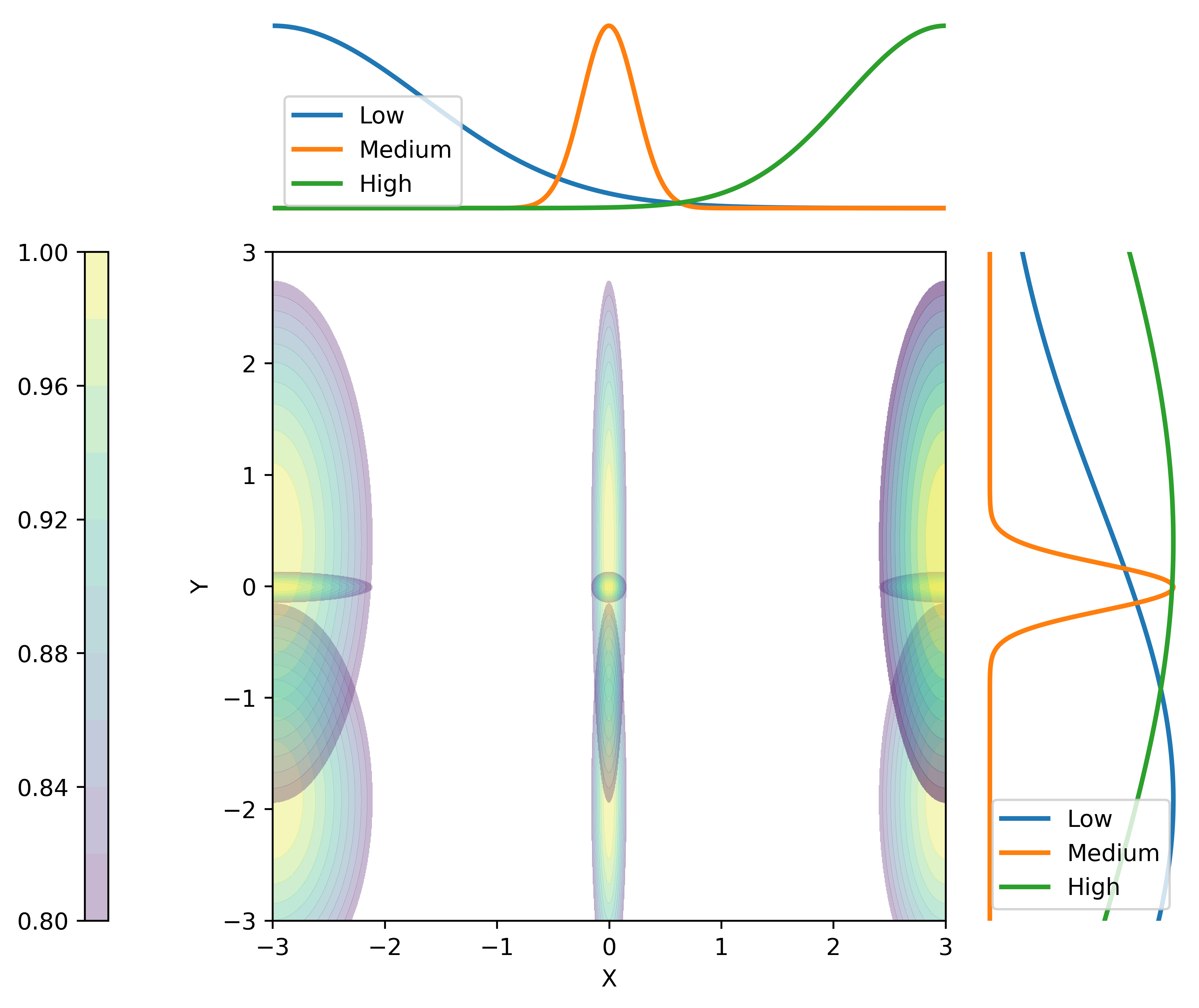}
    \caption{Gaussian membership function activation for the variables space. Greater values in the bidimensional densities show the intersection of the activation rules.}
    \label{fig:gauss_rules}
\end{figure}

Considering now only the ANFIS Gaussian model, the rules activation plot can be visualized in Figure \ref{fig:gauss_rules}. In the marginal graphs, the membership functions are shown individually for each variable of the problem; in each, the three fuzzy rules are colored blue for ``low", orange for ``medium", and green for ``high". Each curve indicates how the input values are mapped into the fuzzy sets, relating to the generated rule.

In turn, the contour plot in the central graph shows the mapping between the X and Y variables under the influence of the three fuzzy rules. The colors represent the degree of rule activation depending on each variable, with the overlapping representing where multiple rules interact. The higher the value is, the more activated a certain rule is. 

\subsection*{Physical Interpretability given by Fuzzy Networks}
\label{sec:interpretability}

The interpretability of ANFIS extends beyond visualizing membership functions and rule activations—it offers a tangible bridge between fuzzy logic constructs and physical quantum behavior. For the H$_2^+$ ion, the Gaussian and Generalized Bell membership functions (MFs) trained by ANFIS encode spatial regions of electron localization. For instance, the Gaussian MFs' centers and variances (e.g., $\mu = \pm 1.2\ \text{\AA }$ along the $x$-axis) align with the equilibrium positions of the two protons, mirroring the expected symmetry of the ion's probability density. The ``high'' membership degree near these coordinates directly correlates with the Born interpretation of |$\Psi|^2$, where electron density peaks at internuclear regions.  

The fuzzy rules further elucidate how ANFIS approximates quantum mechanics. A dominant rule such as:
\begin{quote}
    \textit{``IF \(x\) is near-proton\(_1\) AND \(y\) is low-distance, THEN probability is high''}
\end{quote}
reflects the superposition principle, where constructive interference between atomic orbitals enhances electron density between nuclei. Conversely, rules activating at larger \(x/y\) distances map to low-probability regions, consistent with the exponential decay of wavefunctions in classically forbidden zones.  

Critically, ANFIS reveals \textit{which input combinations dominate predictions}: the stronger activation of rules involving \(x\) (internuclear axis) over \(y\) underscores the 1D symmetry of H$_2^+$'s ground state. This aligns with the Linear Combination of Atomic Orbitals (LCAO) approximation but is derived purely from data-driven fuzzy logic—a novel perspective on orbital hybridization.  

However, ANFIS's rules lack explicit quantum operators (e.g., kinetic energy terms), limiting direct physical analogies. Yet, its ability to distill high-dimensional wavefunction data into a sparse set of interpretable rules provides a heuristic framework for identifying critical spatial features governing electron behavior. For instance, the overlap of ``medium'' membership regions in Figure~\ref{fig:gauss_rules} corresponds to bonding regions, offering a fuzzy-logic analog to molecular orbital theory.  

Therefore, the novel physical insights from ANFIS include:
\begin{itemize}
\item Parameterized Localization: The trained MFs quantify spatial sensitivity — steeper Gaussians (e.g., \(\sigma = 0.5\ \text{\AA}\)) indicate sharper electron confinement, while broader MFs (\(\sigma = 2.0\ \text{\AA}\)) suggest delocalization;  
\item Rule-Based Symmetry: The mirrored MF positions along \(x\) explicitly encode the ion’s symmetry, a feature often obscured in black-box ANN predictions; and  
\item Error-Driven Refinement: ANFIS’s peak-region inaccuracies (Figure~\ref{fig:error}) highlight quantum mechanical nuances (e.g., cusps at nuclei) that fuzzy logic smooths out, guiding future model constraints.  
\end{itemize}

These insights demonstrate ANFIS’s potential as a \textit{physics-discovery tool}, complementing traditional methods by translating wavefunction patterns into human-readable rules. While not replacing Schrödinger-derived solutions, ANFIS offers a complementary lens to interrogate quantum systems through the logic of machine learning.  
While the ANFIS model in this study does not necessarily yield entirely novel physical insights into the dihydrogen ion beyond what is already well-established in quantum mechanics, its interpretability lies in its ability to re-discover and encode these known physical relationships in a transparent and human-understandable manner. The visualization of membership functions and rule activations allows us to see, in a simplified fuzzy logic framework, how the model approximates the complex quantum behavior. This capability is valuable not for groundbreaking new physics in this simple case, but for demonstrating the potential of interpretable AI models like ANFIS to learn and represent physically relevant features directly from data in a way that can be inspected and understood by researchers. For more complex systems where physical intuition might be less straightforward, this interpretability could become crucial for validating model behavior, identifying potentially spurious predictions, and guiding further scientific inquiry based on the model's learned representations.

\section{Conclusion}
This study compared the performance and interpretability of Artificial Neural Networks (ANN) and Adaptive Neuro-Fuzzy Inference Systems (ANFIS) in predicting the wave function of the dihydrogen ion.

Based on the findings, it can be concluded that the Artificial Neural Network outperformed the Adaptive Neuro-Fuzzy Inference System. However, the ANFIS model offers a broader range of possibilities in terms of interpretability, allowing for the visualization of learned membership functions and fuzzy rules. This feature can be crucial in applications where understanding the model is as important as its accuracy.

From the perspective of quantum physics, computational chemistry, and materials science, the probability density prediction is crucial. It not only enables direct access to physical properties but also accelerates quantum simulations and provides a computationally efficient alternative to costly quantum methods. Moreover, an accurate interpretable probability density prediction ensures reliable predictions and might contribute to insights into quantum phenomena and consistency with physical principles.

The results also indicate that the choice of membership function has a significant impact on ANFIS performance. While the Gaussian and Generalized Bell functions yielded similar and superior results compared to the Sigmoid function, combining different membership functions could be explored in future studies to further enhance the model. Additionally, increasing the dataset size proved promising in reducing ANFIS error, suggesting that with a larger dataset, the model could achieve accuracy comparable to that of the ANN. While the use of two independent variables facilitated visualization, future studies could explore more complex systems with additional features to better understand their impact on rule activation.

In practical applications, ANFIS may be preferable in scenarios where interpretability is crucial, such as control systems or modeling complex physical phenomena.
Our findings demonstrate that while Artificial Neural Networks (ANNs) achieve superior predictive accuracy for the H$_2^+$ ion, Adaptive Neuro-Fuzzy Inference Systems (ANFIS) offer a compelling advantage in interpretability alongside remarkable parameter efficiency. The visualized membership functions of the ANFIS model, particularly the Gaussian functions, revealed an encoding of spatial electron localization near the proton positions, directly mirroring Born probability densities. Furthermore, the learned fuzzy rules captured essential quantum principles like superposition and the system's inherent 1D symmetry, offering a novel, data-driven perspective on orbital hybridization aligned with Linear Combination of Atomic Orbitals theory. Analysis of membership function variances and peak prediction errors further provided quantitative insights into electron delocalization trends and limitations in resolving quantum cusps.
Future research can extend ANFIS methodologies to more complex, multi-electron quantum systems and integrate domain-specific constraints, such as kinetic energy terms, to further bridge the gap between data-driven models and the fundamental principles of quantum physics. 
This synergistic approach promises to unlock new avenues for both accurate and insightful quantum simulations.

\section*{Acknowledgements}
The authors thank the support of São Paulo Research Foundation (FAPESP) for the financial support under grants nº 2024/03621-0 and nº 2023/03927-0,
from CNPq project nos. 422069/2023-0, 313301/2025-5, and CNPq - INCT (National Institute of Science and Technology on Materials Informatics) grant no. 371610/2023-0.

\subsection*{Data availability}
The data used in the present work was extracted from %
\cite{physicsinformed_2022}, available at \url{https://github.com/mariosmat/PINN_for_quantum_wavefunction_surfaces}.
\subsection*{Code availability}
The code for the present study is available at the repository: \url{https://github.com/mZaiam/ANFISpy}.

\bibliography{bibliography}
\end{document}